\documentstyle[12pt,epsfig]{article}

\newcommand{\ssh}[1]{#1\!\!\!/}

\newcommand{\half}{{\textstyle\frac{1}{2}}}

\newcommand{\ee}{e^+e^-}
\newcommand{\as}{\alpha_s}
\newcommand{\rs}{\rho_s}

\newcommand{\cl}[1]{{\cal #1}}
\newcommand{\bd}[1]{{\bf #1}}

\newcommand{\sml}[1]{{\mbox{\scriptsize #1}}}
\newcommand{\nml}[1]{{\mbox{#1}}}

\newcommand{\al}{\alpha}
\newcommand{\be}{\beta}
\newcommand{\D}{\Delta}
\newcommand{\de}{\delta}
\newcommand{\e}{\epsilon}
\newcommand{\g}{\gamma}
\renewcommand{\k}{\kappa}

\renewcommand{\r}{\rho}
\newcommand{\s}{\sigma}

\newcommand{\ta}{\theta}

\newcommand{\beq}{\begin{equation}}
\newcommand{\eeq}{\end{equation}}
\newcommand{\bea}{\begin{eqnarray}}
\newcommand{\eea}{\end{eqnarray}}
\newcommand{\nln}{\nonumber\\}
\newcommand{\ind}{\hspace{1.0cm}}

\newcommand{\rf}[1]{(\ref{#1})}
\newcommand{\sect}[1]{\section{#1}\setcounter{equation}{0}
               \hspace{\parindent}\hspace{-0.14cm}}
\newcommand{\sects}[1]{\section*{#1}\setcounter{equation}{0}
               \hspace{\parindent}\hspace{-0.14cm}}
\newcommand{\subsect}[1]{\subsection{#1}
               \hspace{\parindent}\hspace{-0.14cm}}

\newcommand{\jhep}[3]{{\it J.~High Energy Phys.} {\bf #1} (#2) #3}
\newcommand{\npb}[3]{{\it Nucl. Phys.} {\bf B #1} (#2) #3}
\newcommand{\epj}[3]{{\it Eur. Phys.~J.} {\bf #1} (#2) #3}

\newcommand{\plb}[3]{{\it Phys. Lett.} {\bf B #1} (#2) #3}

\newcommand{\prd}[3]{{\it Phys. Rev.} {\bf D #1} (#2) #3}
\newcommand{\prlett}[3]{{\it Phys. Rev. Lett.} {\bf #1} (#2) #3}
\newcommand{\prep}[3]{{\it Phys. Rept.} {\bf #1} (#2) #3}

\newcommand{\zpc}[3]{{\it Zeit. Phys.} {\bf C #1} (#2) #3}

\begin{document}

\begin{titlepage}
\begin{flushright}
Bicocca-FT-01-04 \\
hep-ph/0101323 \\
January 2001
\end{flushright}              
\vspace*{\fill}
\begin{center}
{\Large \bf On the $1/Q$ correction to the $C$-parameter \\[1ex]
at two loops}
\end{center}
\par \vskip 5mm
\begin{center}
        G.E.~Smye\footnote{Research supported by the EU Fourth Framework
                Programme, `Training and Mobitily of Researchers', 
                Network `Quantum Chromodynamics and the Deep Structure of
                Elementary Particles, contract FMRX-CT98-0194 (DG12 - MIHT).}\\
        Dipartimento di Fisica, Universit\`{a} di Milano-Bicocca,\\
        and INFN Sezione di Milano, Italy.\\
\end{center}
\par \vskip 2mm
\begin{center} {\large \bf Abstract} \end{center}
\begin{quote}
We provide an analytical calculation at the two-loop level of the real non-abelian contribution to the leading ($1/Q$) correction to the mean value of the $C$ parameter in $\ee$ annihilation, to complement the existing calculation of the abelian contribution; and we compare the result with the numerical `Milan factor' obtained using the soft approximation. We find agreement with the previous results. The use of the pinch technique to separate the various contributions yields insights into the structure of renormalon-type diagrams in a non-abelian theory.
\end{quote}
\vspace*{\fill}
\end{titlepage}

\sect{Introduction}
The study of event shape variables in $\ee$ annihilation and in deep inelastic scattering continues to provide important insights into the structure of QCD. On the theoretical side, the state-of-the-art perturbative predictions comprise exact next-to-leading order calculations and (for the distributions) a resummation of large logarithms. In addition to the predictions of perturbation theory, large non-perturbative contributions must also be included. These often take the form of power corrections, which are terms behaving as $1/Q^n$ where $Q$ is the hard scale of the process and $n$ is some power.

It has long been known that totally inclusive QCD observables, including for example the total $\ee$ cross-section, possess operator product expansions that may be applied to the calculation of non-perturbative contributions \cite{cond2}. However, no such expansion is available for event shape variables, so some physically reasonable (and experimentally testable) assumptions are required before progress can be made. Either renormalon methods are employed (for a review see \cite{ren3}) with the assumption of ultra-violet dominance \cite{uvdom}, or else assumptions are made concerning the behaviour of the non-perturbative strong coupling in the infra-red, leading to the dispersive approach \cite{dmw}. In practice the two methods give consistent results, but in the calculations that follow the dispersive approach will be adopted.

We consider the resummation of loop insertions into a gluon propagator, which gives rise to a QCD effective charge. (Strictly speaking, this is valid only for fermionic loop insertions, but gluon loops may also be included using `naive non-abelianisation' \cite{nna1,nna2}, or using the 1-loop pinch technique \cite{pt1}-\cite{jay}.) This perturbatively-defined effective charge diverges in the infra-red near the Landau pole, leading to an ill-defined perturbative prediction. But if it is postulated that the full non-perturbative coupling does not have this pole, but rather is well-defined and analytic throughout the complex plane except for a branch cut along the negative real axis, then well-defined predictions can be made. A dispersive representation for this coupling is adopted, in which the dispersive variable plays a role formally similar to that of a gluon mass. The non-perturbative contribution is defined to be the difference between the full prediction and that already included in the fixed-order perturbative result: for event shape means the leading correction is found to be proportional to $\cl{A}_1/Q$, where
\beq
\label{Apdef}
\cl{A}_1 = \frac{C_F}{2\pi}\int_0^\infty d\mu^2\,\mu\delta\as(\mu^2)
\eeq
is the first half-integer moment of the non-perturbative component of the coupling, $\delta\as(\mu^2)$.

It was however quickly pointed out \cite{nase} that the direct application of this method fails to take sufficiently into account the details of the gluon splitting, and that a more sophisticated application is required which treats this splitting correctly. Exact calculations in the abelian limit have been performed for the $1/Q$ corrections to the longitudinal cross section \cite{sigL} and the mean value of the $C$-parameter \cite{dms} in $\ee$ annihilation, while a more general analysis of the most commonly used variables, including non-abelian terms, was performed using the soft approximation in \cite{mf1}-\cite{mf3}. It is found that the effect of non-inclusiveness is simply to enhance the amplitude of the $1/Q$ corrections by a universal `Milan factor', thus leaving the universality pattern unchanged. The extraction of the non-perturbative parameter $\cl{A}_1$ (or, more usually, the related quantity called $\alpha_0$) from a variety of event shape means and distributions is thus a stringent test of the model (as performed for example by \cite{biebel}). It is pleasing to find that universality of the coupling approximately holds.

In the present paper we extend the exact $\ee$ $C$-parameter calculation to include all non-abelian real emission terms, using the 1-loop pinch technique as a convenient tool for the separation of contributions. Virtual contributions are not calculated here, but are included in the same manner as in \cite{mf1}. We recover the Milan factor of \cite{mf1}, which we present in closed form.

The layout of the present paper is as follows: the relevant existing results are recalled in section \ref{sec_ex}, while the calculation of the real non-abelian diagrams is presented in section \ref{sec_calc}. A discussion follows in section \ref{sec_disc}, in which we include the contributions from virtual diagrams and comment on the results.

\sect{Recollection of existing results}
\label{sec_ex}

\subsect{The $C$-parameter}
The $C$ parameter \cite{Cparam} is defined by
\beq
C = \frac{3}{2}\sum_{i,j=1}^{n}\frac{\vert\bd{p}_i\vert\vert\bd{p}_j\vert}{Q^2}\sin^2\ta_{ij}
\eeq
where the sums run over all final-state particles, with $\bd{p}_i$ being the 3-momentum of the $i$th particle in the centre-of-momentum frame of the collision (i.e.~$\sum_i\bd{p}_i=0$), and $\ta_{ij}$ is the angle between $\bd{p}_i$ and $\bd{p}_j$. The hard scale is $Q=\sum_i\vert\bd{p}_i\vert$.

An equivalent frame-independent definition is
\beq
\label{Cdef}
C = 3 - \frac{3}{2}\sum_{i,j=1}^{n}\frac{(p_i\cdot p_j)^2}{(p_i\cdot q)(p_j\cdot q)} \;,
\eeq
where the $p_i$ are the 4-momenta of the $n$ final-state particles, and $q$ is that of the decaying virtual photon, $q=\sum_i p_i$. It is seen that $C=0$ for a pencil-like event, and $C=1$ for a spherical event. In particular, $C$ vanishes at Born level, so the first non-trivial order in perturbation theory is $\cl{O}(\as)$.

At first order in $\as$, let $p_1$, $p_2$ and $k$ be the respective momenta of the outgoing quark, antiquark and gluon. Then $C$ takes the form
\beq
\label{Cmless}
C = 6\frac{(1-x_1)(1-x_2)(1-x_3)}{x_1 x_2 x_3} \;,
\eeq
where $x_{1,2} = 2p_{1,2}\cdot q/q^2$ and $x_3 = 2k\cdot q/q^2$ are the energy fractions carried by the quark, antiquark and gluon respectively, and $\sum_i x_i = 2$ by energy conservation. The leading-order result for the mean value of $C$ is then \cite{Cparam}
\beq
\label{Cpt}
\langle C\rangle = \frac{1}{\s}\int d\s\, C = \frac{C_F}{2\pi}\as(4\pi^2-33)+\cl{O}(\as^2) \;.
\eeq

In order to calculate power corrections in the dispersive approach, we need to consider the gluon to be slightly off-shell. First we use the massive gluon scheme, which treats the gluon decay inclusively, and then we consider in more detail the splitting of the gluon into a quark-antiquark pair or into two gluons.

\subsect{Massive gluon scheme}
Suppose we estimate the $1/Q$ power corrections to the mean value of $C$ using the massive gluon prescription. Although this is known not to be a complete calculation, since it treats the gluon decay inclusively, it is a good first approximation and a necessary starting point for the full calculations below.

If the outgoing gluon is slightly off-shell, with virtuality $k^2=\e Q^2$, the expression for $C$ becomes
\beq
\label{Cmass}
C = 3 - 3\frac{(1-x_3+\e)^2}{x_1 x_2} - 3\frac{(1-x_2-\e)^2}{x_3 x_1} - 3\frac{(1-x_1-\e)^2}{x_2 x_3} - 6\frac{\e^2}{x_3^2} \;.
\eeq

Then the expectation value of $C$ is given to $\cl{O}(\as)$ by
\bea
\label{mgint}
\langle C\rangle_\sml{mg} &=& \frac{1}{\s}\int d\s\,C\nln
&=& \frac{\as}{16\pi N_c}\int_0^{1-\e}dx_1\int_{1-x_1-\e}^{\frac{1-x_1-\e}{1-x_1}}dx_2\,W^{\mu\al}W_{\mu\al}^* C\;,
\eea
where $(-ie)(-ig)W^{\al\mu}$ is the tree level matrix element for the decay of the virtual photon with polarisation index $\mu$ into a quark, an antiquark, and a gluon with polarisation index $\al$, and we implicitly sum over fermion spins. Current conservation then implies that $q_\mu W^{\mu\al} = 0$ and $k_\al W^{\mu\al} = 0$.

In terms of the variables $x_i$ we have
\beq
\label{WW}
W^{\mu\al}W_{\mu\al}^* = 8N_c C_F\left[\frac{(x_1+\e)^2+(x_2+\e)^2}{(1-x_1)(1-x_2)}-\frac{\e}{(1-x_1)^2}-\frac{\e}{(1-x_2)^2}\right]\;,
\eeq
and so from \rf{mgint} we obtain
\beq
\langle C\rangle_\sml{mg} = \frac{C_F}{2\pi}\as\left[4\pi^2-33-12\pi\sqrt{\e}+\cl{O}(\e)\right]\;.
\eeq
Note that in the limit $\e\to 0$ we recover the first-order perturbative result \rf{Cpt}, while the $\sqrt{\e}$ term gives rise to a leading power correction of
\beq
\de\langle C\rangle_\sml{mg} = 12\frac{\cl{A}_1}{Q}\;,
\eeq
where the non-perturbative parameter $\cl{A}_1$ is given by \rf{Apdef}.

In fact we obtain the same result whether we use the massive definition of $C$, \rf{Cmass}, or the massless definition \rf{Cmless} with the mass entering only via the phase space \rf{mgint}. This is not always the case: in some cases the two conventions give different results, for example the longitudinal cross section $\s_L$ in $\ee$ annihilation \cite{br1}-\cite{yubr} or the $C$ parameter in DIS \cite{dw2}. The Milan factor formulation is based on the massless shape variable with massive phase space, so this is the approach we must take when we wish to make comparisons.

\begin{figure}[ht]
\begin{center}
\epsfig{file=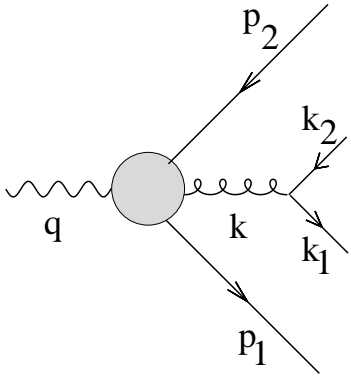, height=5cm}
\caption{\label{absplit}Diagram showing the virtual gluon splitting into a quark
-antiquark pair.}
\end{center}
\end{figure}

\subsect{Abelian gluon splitting}
\label{absec}
Suppose now the virtual gluon splits into a ``secondary'' quark-antiquark pair, with momenta $k_{1,2}$, as in figure \ref{absplit}. The mean value of $C$ now becomes
\bea
\label{Cabint}
\langle C\rangle_\sml{ab} &=& -\int\frac{dk^2}{k^4}\,\frac{T_R n_f}{8\pi N_c}\vert\as(-k^2)\vert^2\int_0^{1-\e}dx_1\int_{1-x_1-\e}^{\frac{1-x_1-\e}{1-x_1}}dx_2\,W^{\mu\al}W_\mu^{*\be}\times\nln
& & \ind\int d\nml{Lips}[k\to k_1,k_2]\,\nml{Tr}[\g_\al\ssh{k}_1\g_\be\ssh{k}_2]C\;.
\eea
Here we have factorised the four-particle phase space into the massive-gluon phase space over $x_1$ and $x_2$, the Lorentz-invariant phase space for the gluon splitting, and an explicit integral over the gluon virtuality $k^2$, which plays the role of dispersive variable. The running coupling $\as(-k^2)$ is generated by making renormalon insertions into both gluon propagators.

For $C$ we take the full expression \rf{Cdef} for four outgoing particles, which we decompose into four pieces:
\beq
\label{Cdecomp}
C = C^{(3)}+C^{(p)}+C^{(m)}+C^{(s)}\;.
\eeq
The first contribution is just $C^{(3)}=3$, the contribution from the first term of \rf{Cdef}, which gives just three times the total cross-section. $C^{(p)}$ contains the remaining ``inclusive'' terms, which are insensitive to the details of the gluon splitting and involve the momenta of the primary fermions only:
\beq
C^{(p)} = -3\frac{(p_1\cdot p_2)^2}{(p_1\cdot q)(p_2\cdot q)}\;.
\eeq
$C^{(s)}$ is the term involving the momenta of the secondary fermions only:
\beq
\label{csec}
C^{(s)} = -3\frac{(k_1\cdot k_2)^2}{(k_1\cdot q)(k_2\cdot q)}\;;
\eeq
while $C^{(m)}$ is the sum of all mixed terms. Using the symmetries of the integral we may replace $k_2$ by $k_1$ and $p_2$ by $p_1$ such that
\beq
C^{(m)} = -12\frac{(p_1\cdot k_1)^2}{(p_1\cdot q)(k_1\cdot q)}\;.
\eeq

We also make manifest the connection with the dispersive approach by writing for each of the contributions
\beq
\langle C^{(i)}\rangle_\sml{ab} = \frac{C_F}{2\pi}\int\frac{dk^2}{k^2}\,\frac{T_R n_f}{3\pi}\vert\as(-k^2)\vert^2\cl{C}^{(i)}_\sml{ab}(k^2/Q^2)\;.
\eeq
The factor $(T_R n_f/3\pi)\vert\as(-k^2)\vert^2$ is none other than $\rs(k^2)$, the spectral function in the large $n_f$ (Abelian) limit. Thus we can compare with the `naive' dispersive treatment, with the functions $\cl{C}^{(i)}$ corresponding to the usual characteristic functions.

These were calculated in \cite{dms}, where the total contribution to the $C$ parameter from abelian gluon splitting was found to be:
\beq
\label{eqcab}
\cl{C}_\sml{ab}(\e) = 4\pi^2-33-\frac{45\pi^3}{32}\sqrt{\e}+\cl{O}(\e)\;.
\eeq

Again we retrieve the standard result \rf{Cpt} in the limit $\e\to 0$, but the coefficient of $\sqrt{\e}$, giving the magnitude of the $1/Q$ power correction, is modified by the full treatment of the gluon splitting. Indeed the exact calculation generates an enhancement of the naively-calculated result by a factor $15\pi^2/128=1.157$.

\begin{figure}[p]
\begin{center}
\epsfig{file=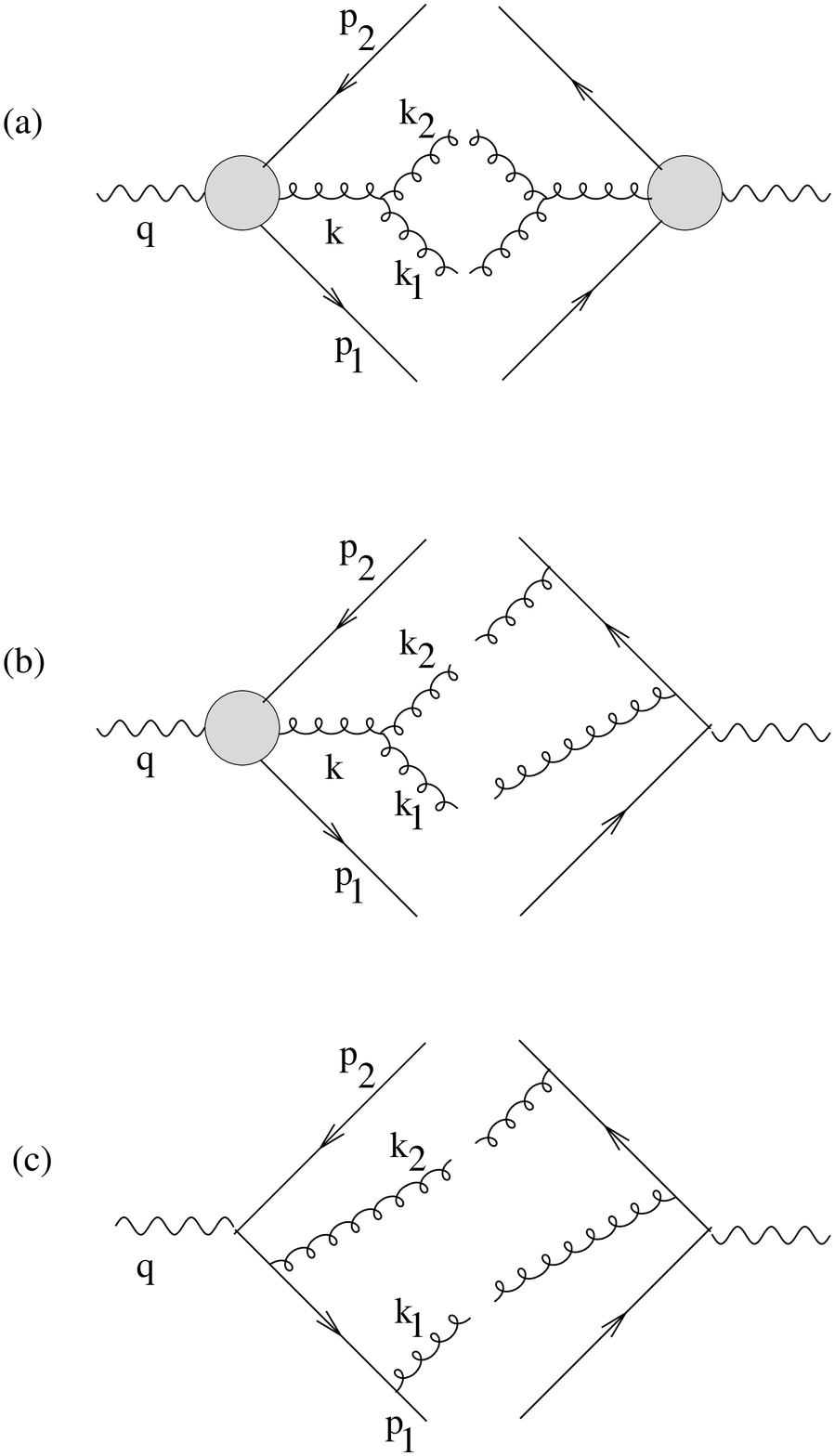, height=16cm}
\caption{\label{nonabspl}Squared Feynman diagrams for the process $\g^*\to q\bar{q}gg$: (a) is the cut loop insertion diagram, with two internal gluon propagators of momentum $k$, (b) and (c) are examples of other diagrams, with one and zero such propagators respectively.}
\end{center}
\end{figure}

\sect{$C$-parameter from $\ee\to q\bar{q}gg$}
\label{sec_calc}
Now let us consider the splitting of the virtual gluon into two real gluons. In order to achieve gauge invariance we must include all tree-level diagrams that contribute to the process $\g^*\to q\bar{q}gg$: we cannot just take the loop insertion diagram shown in figure \ref{nonabspl}(a).

For the purposes of calculation, some prescription must be chosen concerning the gauge-dependent contributions to the diagrams; this usually takes the form of a definite gauge choice. It is helpful here to use the one-loop pinch technique on the squared Feynman diagrams, in order that the cut loop insertion diagram \ref{nonabspl}(a) generates the non-abelian contribution to $\be_0$, behaving analogously to the abelian contribution above, with the remaining diagrams giving additional contributions. Indeed if one recalls the result of inclusive integration over soft gluon emissions given in \cite{mf1}:
\beq
\label{eqinclspl}
\frac{1}{2!}\int_0^1dz\int_0^{2\pi}\frac{d\phi}{2\pi}\left(M_{gg}^2+M_{qq}^2\right) = \frac{1}{m^2(k_\perp^2+m^2)}\left(-\be_0+2C_A\log\frac{k_\perp^2(k_\perp^2+m^2)}{m^4}\right) \;,
\eeq
with $m$ the invariant mass and $k_\perp$ the transverse momentum of the gluon pair, we see that, using the pinch technique, the contribution from figure \ref{nonabspl}(a) generates the term proportional to $\be_0$, while the diagrams like those of figure \ref{nonabspl}(b, c) correspond to the additional logarithmic divergence.

Note that, for the purposes of this calculation, the pinch technique is used simply as a means of reassigning gauge-dependent terms between diagrams, as explained in appendix \ref{apxa}, and no particular significance is attributed to any particular subset of terms.

\subsect{Non-abelian splitting (I): contribution to $\be_0$}
The contribution to $\langle C\rangle$ from the cut loop insertion diagram, figure \ref{nonabspl}(a) using the pinch technique, is given (see appendix \ref{apxa}) by:
\bea
\label{Cniint}
\langle C\rangle_\sml{na1} &=& -\frac{1}{2!}\int\frac{dk^2}{k^4}\,\frac{C_A}{8\pi N_c}\vert\as(-k^2)\vert^2\int_0^{1-\e}dx_1\int_{1-x_1-\e}^{\frac{1-x_1-\e}{1-x_1}}dx_2\,W^{\mu\al}W_\mu^{*\be}\int d\nml{Lips}[k\to k_1,k_2]\nln
& & \ind\times\left\{8(k^2g_{\al\be}-k_\al k_\be)+2(k_{1\al}-k_{2\al})(k_{1\be}-k_{2\be})\right\}C\;.
\eea

This is very similar to equation \rf{Cabint}: apart from the symmetry factor $1/2!$ and the different colour factor, the only difference is in the coefficients of the two terms inside the curly brackets --- recall that in \rf{Cabint} we had:
\beq
\label{trterms}
\nml{Tr}[\g_\al\ssh{k}_1\g_\be\ssh{k}_2] = -2(k^2g_{\al\be}-k_\al k_\be)-2(k_{1\al}-k_{2\al})(k_{1\be}-k_{2\be})\;.
\eeq

Thus the integration proceeds in precisely the same manner as in section \ref{absec}, yielding the result:
\beq
\langle C^{(i)}\rangle_\sml{na1} = -\frac{C_F}{2\pi}\int\frac{dk^2}{k^2}\frac{11C_A}{12\pi}\vert\as(-k^2)\vert^2\cl{C}_\sml{na1}^{(i)}(k^2/Q^2)\;,
\eeq
where the total characteristic function is now
\beq
\label{eqcse}
\cl{C}_\sml{na1}(\e) = 4\pi^2-33-\frac{261\pi^3}{176}\sqrt{\e}+\cl{O}(\e)\;.
\eeq
For completeness the individual components $\cl{C}_\sml{na1}^{(i)}(\e)$ are listed in appendix \ref{apxb}.

The factor $(-11C_A/12\pi)\vert\as(-k^2)\vert^2$ is the `spectral function' for $n_f=0$, which is negative in this case. The running coupling $\as(-k^2)$ is generated by making renormalon insertions into both gluon propagators, as it is possible to do using the one-loop pinch technique \cite{jay}. As before, the characteristic function reproduces the first-order perturbative result \rf{Cpt} in the limit $\e\to 0$, and gives a modification to the naive value of the $1/Q$ correction because of the non-inclusive nature of the observable.

This contribution gives an enhancement of $87\pi^2/704=1.220$ over the `massive gluon' calculation: the reason this is different from that found in the splitting $g\to q\bar{q}$ is because the splittings have different geometry, which in turn is due to the different coefficients in equations \rf{Cniint} and \rf{trterms}.

\subsect{Non-abelian splitting (II): logarithmic divergence}
The diagrams that give rise to the logarithmic term in equation \rf{eqinclspl} are those with only one or no internal gluon propagator, of which examples are given in figure \ref{nonabspl}(b, c). These diagrams are considerably harder to evaluate than the cut loop insertion diagrams, for two reasons. Firstly, they are collinear divergent: an internal quark propagator can be made to go on-shell when one of the real gluons becomes collinear with either the quark or the antiquark. Thus we need a regulator, so we introduce a quark mass $m_q^2$ where
\beq
m_q^2/Q^2 = \D \ll \e
\eeq
and we neglect terms that vanish as $\D\to 0$. The divergent terms must however cancel in the sum of all real and virtual diagrams.

Secondly, the squared matrix element does not as previously factorise into a tensor $W^{\mu\al}W_\mu^{*\be}$ independent of $k_1$ and $k_2$ and a tensor representing the bubble. Thus the integrals become more complicated.

First we consider the 24 diagrams with a single internal gluon propagator, an example of which is given in figure \ref{nonabspl}(b). We write
\beq
\label{eqCna2}
\langle C\rangle_\sml{na2} = -\frac{1}{2!}\int\frac{dk^2}{k^2}\frac{C_A C_F}{16\pi}\as^2\int_0^{1-\e}dx_1\int_{1-x_1-\e}^{\frac{1-x_1-\e}{1-x_1}}dx_2\int d\nml{Lips}[k\to k_1,k_2]\vert M\vert^2 C\;,
\eeq
where the total amplitude contribution from these diagrams evaluated using the pinch technique is
\beq
\left(\half C_A C_F N_c\right) e^2 g^4\frac{\vert M\vert^2}{k^2}\;.
\eeq
We again use the decomposition \rf{Cdecomp} of $C$ into primary, secondary and mixed terms, choosing to write
\beq
\label{eqCna2disp}
\langle C^{(i)}\rangle_\sml{na2} = -\frac{C_F}{2\pi}\int\frac{dk^2}{k^2}\frac{11C_A}{12\pi}\as^2\,\cl{C}_\sml{na2}^{(i)}(k^2/Q^2)\;.
\eeq

The details of the pinch technique require (see appendix \ref{apxa}) that
\beq
\label{eqpin2}
\vert M\vert^2 = \left\{(k_2\!-\!k_1)^\g g^{\al\be}-2k^\al g^{\be\g}+2k^\be g^{\g\al}\right\}\nml{Tr}\left[\ssh{p}_1\g_\al\frac{1}{\ssh{p}_1+\ssh{k}_1}\g_\be\frac{1}{\ssh{p}_1+\ssh{k}}\g^\mu\ssh{p}_2\g_\mu\frac{1}{\ssh{p}_1+\ssh{k}}\g_\g+\cdots\right]
\eeq
where the ellipsis represents 23 similar terms and the factor in curly braces arises from a modified triple-gluon vertex. (The contributions from the remainder of the triple-gluon vertex have been assigned in the pinch technique to other diagrams.)

It is convenient to perform the integral over $k_1$ and $k_2$ in the rest frame of $k$, in which we may write:
\bea
k &=& \sqrt{k^2}(1,0,0,0) \\
k_1 &=& \half\sqrt{k^2}(1,\sin\ta\cos\phi,\sin\ta\sin\phi,\cos\ta) \\
k_2 &=& \half\sqrt{k^2}(1,-\sin\ta\cos\phi,-\sin\ta\sin\phi,-\cos\ta) \\
q &=& (q^0,0,0,q^3) \\
p_1 &=& (p_1^0,0,p_1^2,p_1^3)
\eea
and
\beq
\label{krest2}
d\nml{Lips}[k\to k_1,k_2] = \sin\ta d\ta d\phi/(32\pi^2)\;.
\eeq
It is in performing these integrals over $\ta$ and $\phi$ that the collinear divergence appears: in order to regulate it we substitute
\beq
\frac{1}{\ssh{p}_1+\ssh{k}_1}=\frac{\ssh{p}_1+\ssh{k}_1}{(p_1+k_1)^2} \longrightarrow \frac{\ssh{p}_1+\ssh{k}_1}{(p_1+k_1)^2-m_q^2}=\left.\frac{\ssh{p}_1+\ssh{k}_1}{2p_1\cdot k_1}\right\vert_{p_1^2=m_q^2}\;,
\eeq
and similarly for the other terms. The integral then converges, with the divergences appearing as terms in $\log\D$.

In order to integrate over $x_1$ and $x_2$, it proves convenient to make use of the substitutions 
\bea
\label{usub}
u &=& x_1-x_2 \\
\label{vsub}
v+\frac{\e}{v} &=& 2-x_1-x_2 = x_3 \;.
\eea
After performing the integral over $u$ we find we need to evaluate an integral over $v$,
\beq
\int_{\sqrt{\e}}^1 dv\,f(v,\e)\;,
\eeq
for some complicated function $f$ involving logarithms and dilogarithms of rational functions of $v$ and $\e$. A direct expansion of the integrand in powers of $\e$ is not valid, but may be done by first dividing up the integration range
\beq
\label{eqIeps}
\int_{\sqrt{\e}}^1 dv\,f(v,\e)=\sqrt{\e}\int_1^{1/\k}dy\,f(\sqrt{\e}y,\e)+\int_{\sqrt{\e}/\k}^1dv\,f(v,\e)\;,
\eeq
where $\k$ is a small parameter, $\sqrt{\e}\ll\k\ll 1$, on which the final result does not depend. (In practice we perform the calculation in the limit of small $\k$.)

The second term may now be evaluated by expanding in powers of $\e$. Non-integer powers of $\e$ come only from the lower limit and always appear in the combination $\sqrt{\e}/\k$. In order to avoid having to keep track of terms $\e^m (\k/\sqrt{\e})^n$, where $m$ and $n$ are large, we finally expand the result in $\k$ and neglect all terms that vanish as $\k\to 0$. We obtain
\beq
\label{eqvint}
\int_{\sqrt{\e}/\k}^1dv\,f(v,\e) = a(\log\e)+\frac{\sqrt{\e}}{\k}b(\log\k)+\cl{O}(\e)\;,
\eeq
where $a$ and $b$ are polynomials.

The first term of \rf{eqIeps} may be evaluated by expanding the integrand directly in powers of $\e$. This gives
\beq
\label{eqyint}
\sqrt{\e}\int_1^{1/\k}dy\,f(\sqrt{\e}y,\e)=\sqrt{\e}\left[-\frac{b(\log\k)}{\k}+c\right]+\cl{O}(\e)\;,
\eeq
where the constant $c$ does not depend on $\e$ or $\k$. Adding this to the result \rf{eqvint} gives the total contribution:
\beq
\int_{\sqrt{\e}}^1 dv\,f(v,\e)=a(\log\e)+c\sqrt{\e}+\cl{O}(\e)\;.
\eeq

The individual components $\cl{C}_\sml{na2}^{(i)}(\e)$ are presented in the appendix; the total contribution from the 24 diagrams with a single internal gluon propagator was then found to be:
\beq
\label{eqctr}
\cl{C}_\sml{na2}(\e) = \frac{12}{11}(4\pi^2-33)\log(\e\D)-\frac{1350}{11}+\frac{1440}{11}\zeta(3)-\sqrt{\e}\left(\frac{36\pi^2}{11}\log\D+\frac{144\pi^2}{11}+\nu\right)+\cl{O}(\e)
\eeq
where $\nu\approx-89.556$ is defined in equation \rf{nudef}.

Note that:
\begin{enumerate}
\item The most divergent $\log^3$ and $\log^2$ terms found in the individual components have cancelled in the sum of contributions. There however remain $\log\e$ and $\log\D$ divergences, which will cancel if we sum all real and virtual diagrams. The leading divergence is again proportional to $(4\pi^2-33)$, the leading-order value of $\langle C\rangle$.

\item The coefficient of $\sqrt{\e}$ in \rf{eqctr} is simply the quantity $c$ in equation \rf{eqyint}. Thus an easy way to extract the coefficient of the $1/Q$ correction without performing the full calculation is to evaluate the integral
\bea
\label{eqsoftint}
\cl{I}(\k) &=& \int_1^{1/\k}dy\,\lim_{\e\to 0}f(\sqrt{\e}y,\e) \\
&=& \frac{3}{704}\int_1^{1/\k}\frac{(y^2-1)dy}{2y^2}\int_{-(y-1/y)}^{(y-1/y)}dz\int_{-1}^1 d\cos\ta\int_0^{2\pi}\frac{d\phi}{2\pi}\lim_{\e\to0}\sqrt{\e}\vert M\vert^2 C\nonumber\;,
\eea
which is obtained by combining the definitions \rf{eqCna2} and \rf{eqCna2disp} with the changes of variable \rf{krest2}, \rf{usub} and \rf{vsub}, and by making the substitutions $v=\sqrt{\e}y$ and $u=\sqrt{\e}z$. The coefficient of the $1/Q$ correction is obtained by subtracting off the pole at $\k=0$, just as in \rf{eqyint}. The integrand thus becomes identical with that obtained using soft matrix elements, as in \cite{mf1,mf2}, but the phase-space boundary is more carefully defined in terms of the small arbitrary quantity $\k$.
\end{enumerate}

The final diagrams are those such as figure \ref{nonabspl}(c), diagrams which have no internal gluon propagator. There are 36 such diagrams, 18 of which have colour factor $C_F^2$ and 18 of which have $C_F^2-\half C_F C_A$. Let us choose to write the contributions to the mean value of the $C$ parameter from these diagrams respectively as
\bea
\langle C^{(i)}\rangle_\sml{na3} &=& \frac{C_F}{2\pi}\int\frac{dk^2}{k^2}\frac{11C_F}{6\pi}\as^2\,\cl{C}_\sml{na3}^{(i)}(k^2/Q^2)\nln
\langle C^{(i)}\rangle_\sml{na4} &=& \frac{C_F}{2\pi}\int\frac{dk^2}{k^2}\frac{11(2C_F\!-\!C_A)}{12\pi}\as^2\,\cl{C}_\sml{na4}^{(i)}(k^2/Q^2)\;.
\eea

The integration is performed in the same manner as above. The individual components are listed in appendix \ref{apxb}, and we find evaluate the total characteristic functions to be:
\bea
\label{eqcbx}
\cl{C}_\sml{na3}(\e) &=& -\sqrt{\e}\left(\frac{36\pi^2}{11}\log\D+\frac{36\pi^3}{11}+\nu\right)+\cl{O}(\e) \nln
\cl{C}_\sml{na4}(\e) &=& -\frac{12}{11}(4\pi^2-33)\log\D+\frac{648}{11}-\frac{576}{11}\zeta(3)\nln
& & \ind+\sqrt{\e}\left(\frac{36\pi^2}{11}\log\D+\frac{36\pi^3}{11}+\nu\right)+\cl{O}(\e)
\eea
Again the leading divergence is as expected proportional to $(4\pi^2-33)$.

Now that all the diagrams have been calculated, we are in a position to write down the total real contribution to the $1/Q$ correction to $\langle C\rangle$.

\subsect{Total real contribution}
The total real contribution to the mean value of the $C$-parameter, expanded in powers of $\e$, is, by simple compilation of the results above,
\bea
\label{eqctot}
\langle C\rangle &=& \frac{C_F}{2\pi}\int\frac{dk^2}{k^2}\left[\frac{-\be_0}{4\pi}\vert\as(-k^2)\vert^2\left\{\left(4\pi^2-33\right)-\frac{12\pi\sqrt{\e}}{\be_0}\left(\frac{87\pi^2}{704}\frac{11}{3}C_A-\frac{15\pi^2}{128}\frac{4}{3}T_R n_f\right)\right\}\right.\nln
& & \ind-\frac{11C_A}{12\pi}\as^2\left\{\frac{12}{11}(4\pi^2-33)\log\e-\frac{702}{11}+\frac{864}{11}\zeta(3)-12\pi\sqrt{\e}\frac{12\pi-3\pi^2}{11}\right\}\nln
& & \ind\left.-\frac{2C_F}{\pi}\left\{(4\pi^2-33)\log\D-54+48\zeta(3)\right\}+\cl{O}(\e)\right]\;,\qquad\e=k^2/Q^2\;,
\eea
where the first line comes from the insertions into the gluon propagator, \rf{eqcab} and \rf{eqcse}, and the second and third lines from the remaining diagrams \rf{eqctr} and \rf{eqcbx}.

The factor $(-\be_0/4\pi)\vert\as(-k^2)\vert^2$ is naturally interpreted as the spectral function $\rho_s(k^2)$, generated by making renormalon insertions into the gluon propagators. It is not clear that the same assignment can be made with the remaining contributions, since there is not a simple diagrammatic way of generating the running in diagrams without propagators in which to insert renormalons. If naive non-abelianisation gives the terms in the first line of \rf{eqctot}, the terms on the second and third lines are extra. Nevertheless, the Milan factor is calculated on the assumption that this assignment is valid, and thus we obtain for the real part of the Milan factor
\bea
\label{eqMreal}
\cl{M}_\sml{real} &=& \frac{1}{\be_0}\left\{\frac{11}{3}C_A\left(\frac{12\pi-3\pi^2}{11}+\frac{87\pi^2}{704}\right)-\frac{4}{3}T_R n_f\frac{15\pi^2}{128}\right\}\nln
&=& \frac{3}{64}\frac{(256\pi-35\pi^2)C_A - 10\pi^2 T_R n_f}{11C_A-4T_R n_f}\;.
\eea

\sect{Discussion}
\label{sec_disc}
In the preceding sections we have calculated the mean value of the $C$-parameter in $\ee$ annihilation from 2-loop real emission diagrams as a function of the invariant mass $k^2=\e Q^2$ of the two secondary partons. The coefficient of the $\sqrt{\e}$ term in the series expansion is then assumed to determine the magnitude of the $1/Q$ power correction.

There are a number of points to note here. The first is that virtual contributions have not been calculated exactly. We include them here in the same manner as in \cite{mf1,mf2}, using the soft approximation: 
\beq
\cl{M}_\sml{virt} = -\frac{2\pi(1-\log2)C_A}{\be_0}
\eeq
Details of how this number is extracted from \cite{mf1} are found in appendix \ref{apxc}. Thus we predict the total $1/Q$ correction to the $C$-parameter to be
\beq
\de\langle C\rangle = 12\cl{M}\frac{\cl{A}_1}{Q}\;,
\eeq
where the 2-loop enhancement factor is:
\beq
\cl{M} = \frac{3}{64}\frac{(128\pi+128\pi\log2-35\pi^2)C_A-10\pi^2 T_R n_F}{11C_A-4 T_R n_F}
\eeq
Putting $n_f=3$ yields $\cl{M}={\scriptsize\frac{2}{3}}\pi(1+\log2)-{\scriptsize\frac{5}{24}}\pi^2\approx 1.49$, in agreement with the existing numerical calculations \cite{mf2,blois}. The use of the soft approximation in calculating such enhancement factors is therefore shown to be valid.

It may not at first sight be totally clear exactly how the virtual diagrams, which do not have any phase-space restriction, could contribute to the $1/Q$ correction in the full two-loop approach adopted here. However a term proportional to $\sqrt{\mu^2}/Q$, where $\mu^2$ is a dummy variable in the loop integral, might well appear even without a phase-space restriction if the integrand happens to be of the correct form.

It is also not clear exactly how the spectral function $\rho_s(k^2)$ is generated diagrammatically --- in the large $n_f$ limit this is achieved by quark-loop renormalon insertions into the gluon propagator, which can be extended to gluon-loop insertions using the pinch technique. This reconstructs the correct value for $\be_0$, and is a diagrammatic expression of naive non-abelianisation. But there are left-over contributions to which this does not apply. Whether or not these can truly be said to contribute on the same basis as the renormalon-type contributions is an open question.

\sects{Acknowledgements}
The author would like to thank Mrinal Dasgupta and Lorenzo Magnea, the co-authors of \cite{dms}, and also Yuri Dokshitzer, Giuseppe Marchesini, Gavin Salam, Jay Watson and Bryan Webber, for helpful advice. The author also acknowledges the use of computing facilities at the Cavendish Laboratory, Cambridge, and at CERN.

\setcounter{section}{0}
\renewcommand{\thesection}{\Alph{section}}
\sect{Appendix --- diagram decomposition}
\label{apxa}
We use the pinch technique \cite{pt1}-\cite{pt3} to reallocate certain gauge-dependent terms between diagrams. We first write the diagrams in the simplest gauge for the process, a naive Feynman gauge with external ghost fields to cancel the unphysical degrees of freedom, and then apply pinch technique manipulations and re-assignments. Of course by gauge independence the same results may be obtained with any gauge choice (for a detailed analysis, see \cite{jay}).

The diagrams in figure \ref{nonabspl}(c) are (except for the colour factor) the same as those that appear in QED. With this gauge choice these diagrams become simply:
\beq
\left(C_F^2 N_c\right)e^2 g^4\nml{Tr}\left[\ssh{p}_1\g^\al\frac{1}{\ssh{p}_1+\ssh{k}_1}\g^\be\frac{1}{\ssh{p}_1+\ssh{k}}\g^\mu\ssh{p}_2\g_\mu\frac{1}{\ssh{p}_1+\ssh{k}}\g_\be\frac{1}{\ssh{p}_1+\ssh{k}_1}\g_\al\right]+\cdots
\eeq
where the ellipsis represents the remaining 35 diagrams. We calculate these diagrams just as they are, without further modification.

The diagrams containing a single triple gluon vertex and gluon propagator, such as in figure \ref{nonabspl}(b), are given in this gauge by:
\bea
\lefteqn{\left(\half C_A C_F N_c\right) e^2 g^4\frac{1}{k^2}\left\{(k_2\!-\!k_1)^\g g^{\al\be}-(k\!+\!k_2)^\al g^{\be\g}+(k\!+\!k_1)^\be g^{\g\al}\right\}\times}\nln
& & \ind\nml{Tr}\left[\ssh{p}_1\g_\al\frac{1}{\ssh{p}_1+\ssh{k}_1}\g_\be\frac{1}{\ssh{p}_1+\ssh{k}}\g^\mu\ssh{p}_2\g_\mu\frac{1}{\ssh{p}_1+\ssh{k}}\g_\g+\cdots\right]\;.
\eea
Here we split the triple gluon vertex into two pieces:
\bea
\left\{(k_2\!-\!k_1)^\g g^{\al\be}-(k\!+\!k_2)^\al g^{\be\g}+(k\!+\!k_1)^\be g^{\g\al}\right\} &=& \left\{(k_2\!-\!k_1)^\g g^{\al\be}-2k^\al g^{\be\g}+2k^\be g^{\g\al}\right\}\nln
& & \ind+\left\{k_1^\al g^{\be\g}-k_2^\be g^{\g\al}\right\}\;.
\eea
The contribution from the first term on the right hand side is that used in the calculation of the diagrams, equation \rf{eqpin2}. The contributions from the second term (such contributions are called the `pinched parts' of the diagrams) are manipulated using identities such as:
\beq
k_2^\be\left[\cdots\frac{1}{\ssh{p}_1+\ssh{k}_1}\g_\be\frac{1}{\ssh{p}_1+\ssh{k}}\cdots\right] = \cdots\left(\frac{1}{\ssh{p}_1+\ssh{k}_1}-\frac{1}{\ssh{p}_1+\ssh{k}}\right)\cdots
\eeq
to bring it finally into the form
\beq
\label{trin}
4\left(C_A C_F N_c\right)e^2 g^4\frac{1}{k^4}\left(k^2 g_{\al\be}-k_\al k_\be\right)\nml{Tr}\left[\ssh{p}_1\g^\al\frac{1}{\ssh{p}_1+\ssh{k}}\g^\mu\ssh{p}_2\g_\mu\frac{1}{\ssh{p}_1+\ssh{k}}\g^\be+\cdots\right]\;.
\eeq
The Dirac trace appearing here is identical to that found in the calculation of diagrams \ref{nonabspl}(a): it contains 4 terms in total. This contribution will be added in due course to the conventional loop insertion diagram.

These loop insertion diagrams, figure \ref{nonabspl}(a), are given by
\bea
\lefteqn{\left(C_A C_F N_c\right) e^2 g^4\frac{1}{k^4}\left\{(k_2\!-\!k_1)^\r g^{\al\be}-(k\!+\!k_2)^\al g^{\be\r}+(k\!+\!k_1)^\be g^{\r\al}\right\}\times}\\
& & \left\{(k_2\!-\!k_1)_\s g_{\al\be}^{}-(k\!+\!k_2)_\al g_{\be\s}+(k\!+\!k_1)_\be g_{\s\al}\right\}\nml{Tr}\left[\ssh{p}_1\g_\r\frac{1}{\ssh{p}_1+\ssh{k}}\g^\mu\ssh{p}_2\g_\mu\frac{1}{\ssh{p}_1+\ssh{k}}\g^\s+\cdots\right]\nonumber
\eea
which reduces to 
\bea
\label{glun}
\lefteqn{\left(C_A C_F N_c\right) e^2 g^4\frac{1}{k^4}\left\{4(k^2 g_{\al\be}-k_\al k_\be)+2(k_{1\al}-k_{2\al})(k_{1\be}-k_{2\be})-(k_{1\al}k_{2\be}+k_{2\al}k_{1\be})\right\}\times}\nln
& & \hspace{5cm}\nml{Tr}\left[\ssh{p}_1\g^\al\frac{1}{\ssh{p}_1+\ssh{k}}\g^\mu\ssh{p}_2\g_\mu\frac{1}{\ssh{p}_1+\ssh{k}}\g^\be+\cdots\right]\;.\hspace{3cm}
\eea

The only remaining contribution is from the cut ghost loop, which may be written
\beq
\label{ghon}
\left(C_A C_F N_c\right) e^2 g^4\frac{1}{k^4}\left\{k_{1\al}k_{2\be}+k_{2\al}k_{1\be}\right\}\nml{Tr}\left[\ssh{p}_1\g^\al\frac{1}{\ssh{p}_1+\ssh{k}}\g^\mu\ssh{p}_2\g_\mu\frac{1}{\ssh{p}_1+\ssh{k}}\g^\be+\cdots\right]
\eeq
and thus the total contribution to the cut loop insertion diagram, using this prescription for rearrangement of gauge-dependent terms, is just the sum of \rf{trin}, \rf{glun} and \rf{ghon}:
\bea
\lefteqn{\left(C_A C_F N_c\right) e^2 g^4\frac{1}{k^4}\left\{8(k^2 g_{\al\be}-k_\al k_\be)+2(k_{1\al}-k_{2\al})(k_{1\be}-k_{2\be})\right\}\times}\nln
& & \hspace{2.5cm}\nml{Tr}\left[\ssh{p}_1\g^\al\frac{1}{\ssh{p}_1+\ssh{k}}\g^\mu\ssh{p}_2\g_\mu\frac{1}{\ssh{p}_1+\ssh{k}}\g^\be+\cdots\right]\;.
\eea
It is this combination of terms that is represented by the diagram \ref{nonabspl}(a) and the calculation corresponding to it.

\sect{Appendix --- individual contributions to $\langle C\rangle$}
\label{apxb}
Here are presented the individual contributions to $\langle C\rangle$, denoted $\cl{C}^{(i)}(\e)$ in the text:

\begin{enumerate}
\item In the massive gluon scheme, we find:
\bea
C_\sml{mg}^{(3)}(\e) &=& 3\log^2\e+9\log\e-\pi^2+15+\cl{O}(\e)\nln
C_\sml{mg}^{(p)}(\e) &=& -3\log^2\e-17\log\e+\pi^2-\frac{223}{6}+\cl{O}(\e)\nln
C_\sml{mg}^{(m)}(\e) &=& 8\log\e+4\pi^2-\frac{65}{6}-12\pi\sqrt{\e}+\cl{O}(\e)\nln
C_\sml{mg}^{(s)}(\e) &=& \cl{O}(\e) \;,
\eea
which give the total
\beq
C_\sml{mg}(\e) = 4\pi^2-33-12\pi\sqrt{\e}+\cl{O}(\e) \;.
\eeq

\item For gluon splitting into a secondary $q\bar{q}$ pair, we obtain:
\bea
\cl{C}_\sml{ab}^{(3)}(\e) &=& 3\log^2\e+9\log\e-\pi^2+15+\cl{O}(\e)\nln
\cl{C}_\sml{ab}^{(p)}(\e) &=& -3\log^2\e-17\log\e+\pi^2-\frac{223}{6}+\cl{O}(\e)\nln
\cl{C}_\sml{ab}^{(m)}(\e) &=& 8\log\e+4\pi^2-\frac{65}{6}-\frac{45\pi^3}{32}\sqrt{\e}+\cl{O}(\e)\nln
\cl{C}_\sml{ab}^{(s)}(\e) &=& \cl{O}(\e)\;,
\eea
with total
\beq
\cl{C}_\sml{ab}(\e) = 4\pi^2-33-\frac{45\pi^3}{32}\sqrt{\e}+\cl{O}(\e)\;.
\eeq

\item From the cut loop diagrams such as in figure \ref{nonabspl}(a), we get:
\bea
\cl{C}_\sml{na1}^{(3)}(\e) &=& 3\log^2\e+9\log\e-\pi^2+15+\cl{O}(\e)\nln
\cl{C}_\sml{na1}^{(p)}(\e) &=& -3\log^2\e-17\log\e+\pi^2-\frac{223}{6}+\cl{O}(\e)\nln
\cl{C}_\sml{na1}^{(m)}(\e) &=& 8\log\e+4\pi^2-\frac{65}{6}-\frac{261\pi^3}{176}\sqrt{\e}+\cl{O}(\e)\nln
\cl{C}_\sml{na1}^{(s)}(\e) &=& \cl{O}(\e)\;,
\eea
with total
\beq
\cl{C}_\sml{na1}(\e) = 4\pi^2-33-\frac{261\pi^3}{176}\sqrt{\e}+\cl{O}(\e)\;.
\eeq

\item The diagrams such as in figure \ref{nonabspl}(b) yield
\bea
\cl{C}_\sml{na2}^{(3)}(\e) &=& \frac{12}{11}\log^3\e+\frac{36}{11}\log\D\log^2\e+\frac{18}{11}\log^2\D\log\e+\frac{54}{11}\log^2\e\nln
& & +\frac{108}{11}\log\D\log\e+\frac{36}{11}\log^2\D-\frac{30\pi^2-261}{11}\log\e-\frac{12\pi^2-171}{11}\log\D\nln
& & -\frac{39\pi^2}{11}+\frac{477}{11}-\frac{270}{11}\zeta(3)+\frac{12\pi^2}{11}\log2-\frac{24}{11}\log^3 2+\frac{144}{11}\nml{Li}_3(\half)+\cl{O}(\e)\nln
\cl{C}_\sml{na2}^{(p)}(\e) &=& -\frac{12}{11}\log^3\e-\frac{36}{11}\log\D\log^2\e-\frac{18}{11}\log^2\D\log\e-\frac{102}{11}\log^2\e\nln
& & -\frac{204}{11}\log\D\log\e-\frac{54}{11}\log^2\D+\frac{30\pi^2-507}{11}\log\e+\frac{12\pi^2-409}{11}\log\D\nln
& & +7\pi^2-\frac{1042}{11}+\frac{270}{11}\zeta(3)-\frac{12\pi^2}{11}\log2+\frac{24}{11}\log^3 2-\frac{144}{11}\nml{Li}_3(\half)+\cl{O}(\e)\nln
\cl{C}_\sml{na2}^{(m)}(\e) &=& \frac{48}{11}\log^2\e+\frac{96}{11}\log\D\log\e+\frac{18}{11}\log^2\D+\frac{48\pi^2-150}{11}\log\e\nln
& & +\frac{48\pi^2-158}{11}\log\D-\frac{38\pi^2}{11}-\frac{785}{11}+\frac{1440}{11}\zeta(3)\nln
& & -\sqrt{\e}\left(\frac{36\pi^2}{11}\log\D+\frac{144\pi^2}{11}+\nu\right)+\cl{O}(\e)\nln
\cl{C}_\sml{na2}^{(s)}(\e) &=& \cl{O}(\e)\;,
\eea
which have total
\bea
\cl{C}_\sml{na2}(\e) &=& \frac{12}{11}(4\pi^2-33)\log(\e\D)-\frac{1350}{11}+\frac{1440}{11}\zeta(3)\nln
& & \ind-\sqrt{\e}\left(\frac{36\pi^2}{11}\log\D+\frac{144\pi^2}{11}+\nu\right)+\cl{O}(\e)\;.
\eea
The number $\nu\approx-89.556$ is given by the expression
\beq
\label{nudef}
\nu = \frac{288\pi}{11}\int_{\pi/6}^{5\pi/6}\log\sin\frac{\ta}{2}d\ta \;-\; \frac{432\pi}{11}\int_{\pi/3}^{2\pi/3}\log\sin\frac{\ta}{2}d\ta \;-\; \frac{96\pi^2\log 2}{11}\;.
\eeq

\item The diagrams such as that of figure \ref{nonabspl}(c) which have colour factor $C_F^2$ give the contributions:
\bea
\cl{C}_\sml{na3}^{(3)}(\e) &=& -\frac{18}{11}\log^2\D\log\e-\frac{18}{11}\log^2\D+\frac{9}{22}\log\e+\frac{24\pi^2-9}{22}\log\D\nln
& & \ind+\frac{6\pi^2}{11}-\frac{9}{22}-\frac{108}{11}\zeta(3)+\cl{O}(\e)\nln
\cl{C}_\sml{na3}^{(p)}(\e) &=& \frac{18}{11}\log^2\D\log\e+\frac{48}{11}\log^2\D-\frac{73}{22}\log\e-\frac{24\pi^2-73}{22}\log\D\nln
& & \ind-\frac{16\pi^2}{11}-\frac{5}{22}+\frac{108}{11}\zeta(3)+\cl{O}(\e)\nln
\cl{C}_\sml{na3}^{(m)}(\e) &=& -\frac{30}{11}\log^2\D+\frac{32}{11}\log\e-\frac{32}{11}\log\D+\frac{10\pi^2}{11}+\frac{7}{11}\nln
& & \ind-\sqrt{\e}\left(\frac{36\pi^2}{11}\log\D+\frac{36\pi^3}{11}+\nu\right)+\cl{O}(\e)\nln
\cl{C}_\sml{na3}^{(s)}(\e) &=& \cl{O}(\e)\;,
\eea
which have total
\beq
\cl{C}_\sml{na3}(\e) = -\sqrt{\e}\left(\frac{36\pi^2}{11}\log\D+\frac{36\pi^3}{11}+\nu\right)+\cl{O}(\e)\;.
\eeq

\item And finally those diagrams with colour factor $C_F^2-\half C_A C_F$ provide:
\bea
\cl{C}_\sml{na4}^{(3)}(\e) &=& -\frac{36}{11}\log\D\log^2\e-\frac{18}{11}\log^2\D\log\e-\frac{108}{11}\log\D\log\e-\frac{36}{11}\log^2\D\nln
& & +\frac{36\pi^2-135}{11}\log\e+\frac{12\pi^2-171}{11}\log\D+\frac{48\pi^2}{11}-\frac{711}{22}+\frac{252}{11}\zeta(3)+\cl{O}(\e)\nln
\cl{C}_\sml{na4}^{(p)}(\e) &=& \frac{36}{11}\log\D\log^2\e+\frac{18}{11}\log^2\D\log\e+\frac{204}{11}\log\D\log\e+\frac{54}{11}\log^2\D\nln
& & -\frac{36\pi^2-219}{11}\log\e-\frac{12\pi^2-409}{11}\log\D-\frac{86\pi^2}{11}+\frac{1487}{22}-\frac{252}{11}\zeta(3)+\cl{O}(\e)\nln
\cl{C}_\sml{na4}^{(m)}(\e) &=& -\frac{96}{11}\log\D\log\e-\frac{18}{11}\log^2\D-\frac{84}{11}\log\e-\frac{48\pi^2-158}{11}\log\D\nln
& & +\frac{38\pi^2}{11}+\frac{260}{11}-\frac{576}{11}\zeta(3)+\sqrt{\e}\left(\frac{36\pi^2}{11}\log\D+\frac{36\pi^3}{11}+\nu\right)+\cl{O}(\e)\nln
\cl{C}_\sml{na4}^{(s)}(\e) &=& \cl{O}(\e)\;,
\eea
with total
\bea
\cl{C}_\sml{na4}(\e) &=& \frac{396-48\pi^2}{11}\log\D+\frac{648-576\zeta(3)}{11}\nln
& & \ind+\sqrt{\e}\left(\frac{36\pi^2}{11}\log\D+\frac{36\pi^3}{11}+\nu\right)+\cl{O}(\e)\;.
\eea
\end{enumerate}

\sect{Appendix --- extraction of virtual contribution}
\label{apxc}
In order to compare our real-emission result \rf{eqMreal} with the Milan factor we have to note that the latter as derived in \cite{mf1,mf2} includes a contribution from virtual diagrams as well as real diagrams. This is a part of the inclusive piece $r_\sml{incl}=c^{(i)} C_A/\be_0$, where from equation (3.8) of \cite{mf1} we have
\beq
\label{eqmfpaper}
c^{(i)} = \int_0^\infty dx\frac{2x}{1+x^2}\left(\sqrt{1+x^2}-x\right)\log\left[x^2(1+x^2)\right] = 8-4\log2-2\pi(1-\log2)\;.
\eeq
Here the contribution involving $\sqrt{1+x^2}$ is the contribution from real emission diagrams, while the virtual contribution is represented by the $-x$ term. The combined integral converges at large $x$.

In the full calculation (i.e.~without using the soft approximation) the real and virtual contributions represented here arise from integrals of the form:
\beq
\cl{K}_\sml{re/vi} = \int_0^1 d\xi\, h_\sml{re/vi}(\xi,\e)\;,
\eeq
where the kinematical integration variable is $\xi=k_\perp/Q$. Using \rf{eqinclspl} and the argument of \cite{mf1} we have in the soft (small $\xi$) region that
\bea
h_\sml{re}(\xi,\e) \sim \frac{2\xi}{\xi^2+\e}\sqrt{\xi^2+\e}\log\frac{\xi^2(\xi^2+\e)}{\e^2} \nln
h_\sml{vi}(\xi,\e) \sim \frac{2\xi}{\xi^2+\e}(-\xi)\log\frac{\xi^2(\xi^2+\e)}{\e^2}\;.
\eea
It is therefore clearly not permissible to evaluate the integrals $\cl{K}_\sml{re/vi}$ by a naive expansion of the integrand in powers of $\e$. Instead we divide up the integration region according to
\beq
\cl{K}_\sml{re/vi} = \sqrt{\e}\int_0^{1/\k}dx\,h_\sml{re/vi}(\sqrt{\e}x,\e)+\int_{\sqrt{\e}/\k}^1d\xi\, h_\sml{re/vi}(\xi,\e)\;,
\eeq
where the total cannot depend on $\k$.

Neglecting terms that vanish as $\k\to0$, the first term on the right-hand side gives
\bea
\label{eqsep}
\int_0^{1/\k}dx\frac{2x}{1+x^2}\sqrt{1+x^2}\log\left[x^2(1+x^2)\right] &=& -\frac{8}{\k}(1-\log\k)+8-4\log2\;,\nln
-\int_0^{1/\k}dx\frac{2x^2}{1+x^2}\log\left[x^2(1+x^2)\right] &=& \frac{8}{\k}(1-\log\k)-2\pi(1-\log2)\;,\hspace{0.5cm}
\eea
while in the second the $\sqrt{\e}$ terms can come only from the lower limit, and thus appear as $\sqrt{\e}/\k$. The $1/\k$ terms in \rf{eqsep} are therefore exactly cancelled, leaving the total $\sqrt{\e}$ contributions from real and virtual parts respectively as
\beq
c_\sml{real}^{(i)} = 8-4\log2 \hspace{1.5cm} c_\sml{virt}^{(i)} = -2\pi(1-\log2)\;.
\eeq

\end{document}